\documentclass[preprint]{vgtc}               

\ifpdf
  \pdfoutput=1\relax                   
  \usepackage{graphicx}                
  \DeclareGraphicsExtensions{.pdf,.png,.jpg,.jpeg} 
\else
  \usepackage{graphicx}                
  \DeclareGraphicsExtensions{.eps}     
\fi%

\graphicspath{{figures/}{pictures/}{images/}{./}} 

\usepackage{microtype}                 
\PassOptionsToPackage{warn}{textcomp}  
\usepackage{textcomp}                  
\usepackage{mathptmx}                  
\usepackage{times}                     
\usepackage{cite}                      
\usepackage{tabu}                      
\usepackage{booktabs}                  

\usepackage{soul}
\usepackage[color=yellow]{todonotes}
\usepackage{balance}
\usepackage{url}

\usepackage{changepage}

\usepackage{adjustbox}
\usepackage{ragged2e}
\usepackage{booktabs}
\usepackage{multicol}
\usepackage{colortbl}

\definecolor{CindyX}{RGB}{232, 125, 114}

\onlineid{0}

\vgtccategory{Research}

\vgtcinsertpkg

\preprinttext{To appear in the Visualization for Social Good Workshop at IEEE VIS 2021.}

\title{Making the Invisible Visible: Risks and Benefits of Disclosing Metadata in Visualization}

\author{Alyxander Burns\thanks{e-mail: alyxanderbur@cs.umass.edu}\\ %
        \scriptsize UMass Amherst %
\and Thai On \\%
     \scriptsize UMass Amherst %
\and Christiana Lee \\ %
     \scriptsize UMass Amherst %
\and Rachel Shapiro \\ %
     \scriptsize Northwestern %
\and Cindy Xiong \\ %
     \parbox{1.4in}{\scriptsize \centering UMass Amherst }
\and Narges Mahyar\\ %
     \scriptsize UMass Amherst %
     }

\abstract{
Accompanying a data visualization with metadata may benefit readers by facilitating content understanding, strengthening trust, and providing accountability. However, providing this kind of information may also have negative, unintended consequences, such as biasing readers' interpretations, a loss of trust as a result of too much transparency, and the possibility of opening visualization creators with minoritized identities up to undeserved critique. To help future visualization researchers and practitioners decide what kinds of metadata to include, we discuss some of the potential benefits and risks of disclosing five kinds of metadata: metadata about the source of the underlying data; the cleaning and processing conducted; the marks, channels, and other design elements used; the people who directly created the visualization; and the people for whom the visualization was created. We conclude by proposing a few open research questions related to how to communicate metadata about visualizations.

} 

\CCScatlist{
  \CCScatTwelve{Human-centered computing}{Visu\-al\-iza\-tion}{Visu\-al\-iza\-tion theory, concepts and paradigms}{}
}

\begin{document}


\firstsection{Introduction}

\maketitle

In this paper, we discuss the pros and cons of accompanying a data visualization with different pieces of \textbf{metadata} which we define as:
\textit{information not directly represented in a visualization that provides background on the source of the data, the transformations applied to the data, the visualization elements, its purpose, the people involved in its creation, and its intended audience.}
This definition of metadata is broad and encompasses the kind of information often considered metadata as well as paradata (traditionally, data about methodology) and information that may not traditionally be thought of as ``data'' but are nonetheless important for establishing the social, cultural, or historical circumstances in which a visualization was made (e.g., authorial statements of positionality which elaborate on identities and experiences that inform how the authors relate to knowledge). Our definition of metadata, therefore, may also encompass the narratives that a visualization is involved in. Although narratives and traditional data-like metadata are different, we chose to group them together in this work in order to consider the broadest range of additional information which could be provided to readers.

Accompanying a data visualization with metadata makes the intentions and decisions of the creators transparent to readers and promotes understanding of the history, culture, and context which lead to the visualization's creation. 
These effects, in turn, interrogate and challenge societal power \cite{dork2013critical, dataFeminism}, and help to democratize data science by putting the information into the hands of more people.
Further, it can provide practical benefits for readers. For example, consider a visualization of development plans for a city. Residents of that area have insider knowledge about the socio-political landscape of the community. With this information, they may be able to conclude, for example, that poor families will be disproportionately impacted by the proposed changes. This kind of expert information is not held by others who are less familiar with the area and while both readers may be able to use the visualization, important nuance may still be lost when expert or personal knowledge is unavailable.

Although providing metadata might help readers make more informed judgements and learn how to understand various elements of a visualization, deciding what to provide and how to do so can be challenging. These decisions require carefully balancing the possible benefits against potential disadvantages, such as the tension between transparency and trust. Providing any kind of metadata increases transparency which can help the general public understand who and what was involved in the final result \cite{elliott2014science} and build trust \cite{horvath2016transparency}. Trust, in turn, can increase how the much a visualization will be used \cite{mayr2019trust, hullman2011visualization}.
However, it is important to carefully decide which and how much information to disclose because the relationship between trust and transparency is not linear -- increasing transparency helps trust only to a point, after which increasing transparency \textit{decreases} trust \cite{horvath2016transparency, kizilcec2016much}.
Additionally, providing too much metadata might bias a reader's interpretation of the visualization, overwhelm them, or expose visualization creators to undue critique or distrust as a result of socio-political biases such as racism, sexism, and xenophobia. 

We draw on existing literature and our experiences as data visualization researchers to consider two broad groups of metadata 
-- metadata about the visualization pipeline and metadata about the people who contributed to the creation of visualizations. 
There are three stages in the pipeline from ``raw'' data to a published visualization: (1) the collection of the original/``raw'' data, (2) the transformations applied to the data (e.g., cleaning, aggregation), and (3) the visualization that is produced \cite{pang1996pipeline}.
Analogously, there are two groups of people involved in the creation of a visualization: (1) the creators who directly change the design and (2) the intended audience whose needs indirectly inform the design choices made.

Our approach is inspired by existing scholarship that challenges academic communities to do more good by making the invisible visible (e.g., \cite{dork2013critical, dataFeminism, correll2019ethical}).
The purpose of this paper is not to exhaustively describe the kinds of the metadata that \textit{could} be provided alongside a visualization or to assert what information \textit{should} be included. Instead, we want to open a discussion within the visualization community about what kinds of information will best support readers under different circumstances and what needs to be done to combat barriers that make doing so difficult.

\section{Background}

\noindent\textbf{Information about data.}
Researchers have long advocated for collecting and disseminating information about data. For example, data analysts have been interested in the lineage and history of data to establish accuracy, replicate processes, and conduct meta-analyses, among other aims  (e.g, \cite{buneman2001and, simmhan2005survey, ragan2015characterizing}). Feminist scholars have emphasized the importance of understanding the social and historical contexts of data to challenge hegemonic power structures (e.g., \cite{dataFeminism,  haraway1988situated, gebru2018datasheets}).
Collecting and disseminating information about data has been argued to be beneficial for prospective users by, for example, helping them interpret data  \cite{simmhan2005survey} and judge whether the application of a dataset is appropriate  \cite{gebru2018datasheets}.
The practice can also be helpful to the research community as a whole for enabling replication \cite{haroz2018open, battle2018evaluating} and conducting meta-analyses \cite{xu2020survey}.

\vspace{4pt}\noindent\textbf{Information about people.} While the typical models of metadata in data science capture many aspects of the history of the data and the transformations applied to it, they do not often include information about the people involved in the process. 
However, all knowledge is reflective of the perspectives of the knower \cite{haraway1988situated}, meaning that one's identities and experiences both frame and limit what knowledge they create \cite{dataFeminism, lee2021viral}. Past work in the visualization community has also demonstrated that a person's personal knowledge and experience is highly related to what they will perceive as useful \cite{peck2019data}, believe to be too difficult to understand \cite{kennedy2016engaging}, and what they think others will see in the visualization \cite{xiong2019curse}. Thus, understanding people and their positionality is an integral part of understanding the data and visualization they created. 

\vspace{4pt}\noindent\textbf{Instructional information.} One could also consider including metadata about how to read a visualization. A reader's interpretation of a visualization may be affected by their own experience, knowledge, and perspectives (e.g., \cite{peck2019data, hullman2011visualization, kennedy2016engaging}). Further, rhetorical choices made by the visualization designer could make some interpretations more or less salient \cite{hullman2011visualization}. Therefore, it may be helpful to include ``instructions'' for decoding a visualization to minimize confusion and misunderstanding, especially when a visualization is complex or when the rhetorical choices are novel.
For example, providing instructions and tips in the form of cheat sheets \cite{wang2020cheat} or slide shows \cite{wang2019narvis} have been shown to help people who are new to visualization draw conclusions from unfamiliar data visualizations.

\vspace{4pt}\noindent\textbf{Challenges to providing information.} Unfortunately, while it might be beneficial to provide lots of metadata to readers, it is critical do so without overwhelming them. We need to carefully select pieces of information to include. 
Past research concluded that providing metadata like data provenance can be a tool for signalling transparency and trustworthiness to end-users \cite{hullman2011visualization}. This is desirable because trust ``increases the likelihood that viewers believe what they see” \cite{hullman2011visualization}.
While increasing transparency is often associated with increased trust (e.g., as in \cite{munzel2016assisting}), it has nonetheless been observed that this relationship does not always hold.
Instead, transparency increases trust to a critical point, at which time increasing transparency further decreases trust --- especially when expectations are broken \cite{horvath2016transparency, kizilcec2016much}.
This erosion of trust is theorized to occur because too much explanation confuses readers and directs their attention away from the explanation and toward the unexpected outcome \cite{kizilcec2016much}.
Further, there may be other negative effects for increasing transparency depending on what kind of information is disclosed (such as harassment sometimes faced by researchers when their names are made public \cite{lewandowsky2016research}). 
The trade-off between transparency and trust makes deciding what metadata to provide a challenging task. 
In this paper, we outline several kinds of metadata relevant to data visualization that might be provided to the public, what benefits that may bring, and what challenges remain.

\section{Types of Metadata}
In this section, we discuss the pros and cons of two broad groups of metadata: metadata about the visualization pipeline from ``raw'' data to final visualization and metadata about the people behind the pipeline. 
We do not wish to suggest that there is a distinct separation between pipeline and people -- there is, in fact, quite the opposite because data are necessarily the result of human decisions and processes.
However, we decided to divide these two groups in order to focus on what kinds of information we could provide about each and because they offer different benefits and challenges.

We identified the two groups based on existing research including Gebru et al.'s ``Datasheets for Datasets'' \cite{gebru2018datasheets} and Krause's notion of ``Data Biographies'' \cite{dataBios}.
Gebru et al. \cite{gebru2018datasheets} divided metadata into seven thematic categories based on important components of a dataset's lifecycle, where five categories are encompassed within our section about the visualization pipeline (Collection Process, Preprocessing/Cleaning/Labeling, Maintenance, Composition, and Uses) and the remaining two fit within our category about people (Motivation and Distribution).
Krause \cite{dataBios} approached the categorization differently by creating five groupings based on the central questions of When, Where, Why, How, and Who. Of these categories, When, Where and How are covered by our section about the visualization pipeline and Who and Why fall within our section on people.

Previously, we discussed that providing any kind of truthful information increases transparency and trust up to a threshold, after which disclosing more may decrease trust. Therefore, in the remainder of this section, we will only mention this relationship when there are novel considerations.

\subsection{Metadata about the Visualization Pipeline}
\label{section:Data}
We consider what metadata could be disclosed to the readers regarding three stages of the data visualization pipeline: data sourcing, data transformation, and data visualization. We primarily focus on the transformation and visualization components because existing work has extensively researched what should be collected and disseminated about the data sourcing step \cite{gebru2018datasheets}.

\subsubsection{Data Source}
Documenting the data collection process itself is a critical component of documenting a dataset and is covered in depth in \cite{gebru2018datasheets}. However, describing which dataset was used in a visualization may still be an important piece of metadata to include. In fact, this practice is already an important part of the visualization style guides for some organizations (e.g., the Urban Institute \cite{urbanInstitute}).

\vspace{4pt}\noindent\textbf{Pros:} 
Providing the name of (or, ideally, a reference link to) the original dataset, may allow readers to replicate the visualization or conduct their own analyses on the data. While not all readers will have the time or expertise to do further analysis, they still might use this information to judge the usefulness of the visualization if they are able to discern its collectors. 
Evidence of using the source to judge usefulness was observed in past work, though the source was of the visualization, not of the underlying dataset \cite{peck2019data}. 

\vspace{4pt}\noindent\textbf{Cons:} However, it may be difficult to clearly describe the source of data without also disclosing who collected it, which may have negative impacts on privacy and trust if, for example, a dataset is associated with a particular individual or organization that readers already distrust. 

\subsubsection{Cleaning and Processing}
However, disclosing \textit{only} the dataset without also describing which transformations were applied to the data may lead to a loss of trust from data analysis-savvy readers who conduct their own analyses. This is because the readers could make different data analysis choices and come to different (or even contradictory) results \cite{silberzahn2018many}.
Data processing is the middle-stage between the data obtained from the dataset and the visualization. It is important to consider what one could disclose about this stage because the methods chosen can introduce uncertainty into the data in ways which are not immediately apparent in the visualization \cite{pang1996pipeline, hullman2011visualization}.
For example, the choice to keep or filter out outliers may change the mean and variation of the data. This, in turn, will change the visualization in ways which may not be apparent to the reader.
Within this stage, one could also consider disclosing whether the data will be cleaned or processed again in the advent of updates to the underlying dataset. This kind of information might be particularly important in domains like public health where the recency of data is critical because there can be consequences for making a decision with out-of-date data.

\vspace{4pt}\noindent\textbf{Pros:} Disclosing one's data cleaning and processing methods make the visualization more credible \cite{marcus2015credibility} and replicable, especially if the methods are sufficiently detailed such that another person could follow the steps to recreate the values represented in the final visualization.
Further, this information may also allow readers to independently extract insights that can improve their own data cleaning methods in the future, as these insights ``cannot be ensured by merely observing the results from data analytics''\cite{choi2020meaning}.

\vspace{4pt}\noindent\textbf{Cons:} On the other hand, being transparent about one’s methods of data cleaning and processing might open the creators up to undue or pedantic critique, especially if the creators belong to minoritized groups (e.g., because minoritized individuals are not always believed even when they are experts \cite{lanius2015fact, fricker2007epistemic}) . 
Furthermore, while pressure to include how the data were processed might push some creators to select their methods more carefully if they know it will be public information, any diversion from the methodological norm might incur the wrath of the crowd and result in creator harassment or a loss of trust in the visualization -- even if the methods are sound.
Finally, it may be deceptively difficult to explain methodology in a way that is understandable for the general public, especially when analysis techniques are complicated or counter-intuitive \cite{rowan1991simple}. Science communication techniques might provide guidance on this front, but explanation remains a challenge.

\subsubsection{Visual Encoding: Explaining Perceptual Challenges}
The final stage in the visualization pipeline is visualizing the data. Because this space is less explored in existing literature, we will discuss two different options for this stage.

One possible piece of information one could provide is an explanation of problems that the reader might face while trying to decode the marks, channels, or other design elements used in the visualization.
For example, log scales have been shown to be difficult for the general public to understand \cite{romano2020scale}. Offering information that explains the log scale or simply points out its existence might help readers better understand the scale and the visualization as a whole.

\vspace{4pt}\noindent\textbf{Pros:}
Describing the potential difficulties of a visualization's encoding may help readers understand what problems exist so that they might try to avoid them. Existing work has shown that when flaws in visualizations are pointed out to readers, they are able to identify similar errors in other charts more often in the future \cite{2020-visualint}. This suggests that explaining what may go wrong with a visualization may also help readers learn over time what to look out for. 
Finally, it is possible that visualization creators will also learn how to create better designs over time by thinking critically about the negative impacts their design choices. Reflection is already an integral part of generating insights for visualization design studies \cite{meyer2018reflection} and it stands to reason that design insights could result from asking designers to reflect on their designs more generally.

\vspace{4pt}\noindent\textbf{Cons:} Nevertheless, when a reader is made aware of potential pitfalls in a visualization, it is possible that it becomes irredeemable in their opinion. The relationship between trust and use after a system is known to be flawed seems to not be very well understood in data visualization, but is an ongoing research direction in the related field of human-robot interaction where existing work suggests that the impact of known flaws on the user's behavior changes with the kind of task they are conducting (e.g., in \cite{salem2015would}).
Additionally, telling the reader that there are potential issues with a design begs the question: Why wasn't a different design choice made?
Every design choice involves trade-offs that may be clear to the visualization creator but not to the reader. Finding ways to communicate what choice was made and why could bring the readers and creators onto the same page regarding the visualization design, but justification is particularly difficult when readers do not posess sufficient knowledge of the alternatives.
Finally, explaining to the reader that they may make a mistake understanding an aspect of a visualization takes up space in the metadata that could be used to include other useful information and does not necessarily prevent the reader from making that mistake. For example, past work on truncated y-axes found that participants using visualizations that explicitly pointed out truncated axes made similar errors to participants using designs that did not point out the potential problem \cite{correll2020truncating}.

\subsubsection{Visual Encoding: How to read the visualization}
Finally, one could also consider providing information that helps the reader make sense of a visualization. 
Two possible ways of doing this are (1) directly telling the reader what the intended message of the visualization is and (2) providing a tutorial describing how to ``decode'' the encoding. 
For example, when explaining a scatterplot, one could directly report that sugar intake and the number of cavities have a positive correlation or break down the encoding by describing what each point represents and what the axes mean without providing a specific conclusion that could or should be drawn. 
These two techniques have overlapping benefits but different challenges.

\vspace{4pt}\noindent\textbf{Pros:} Directly providing information about what is being shown through either of the methods described may primarily benefit readers who are less experienced in decoding visualizations.
Past work has shown that generating slideshows and ``cheat sheets,'' which explained the encodings of complicated data visualizations, helped readers without a background in visualization successfully extract information from the charts \cite{wang2019narvis, wang2020cheat}.
It is also possible that this type of information may motivate people to engage with and try to understand complex or unfamiliar looking visualizations by boosting self-efficacy, a factor known to impact engagement \cite{kennedy2016engaging}.
Finally, one possible benefit of providing a tutorial (which is not necessarily afforded by describing the intended message) is that readers may learn how to decode similar visualizations in the future.

\vspace{4pt}\noindent\textbf{Cons:} However, there are also drawbacks to providing either the intended message or a tutorial.
For instance, being explicit about the intended message might bias attention and interpretation, as well as change the reader's behavior. Pointing out a feature or conclusion to a visualization reader can change what they pay attention to \cite{xiong2019curse}.
For example, the content of captions can bias reader's attention toward specific points instead of other salient ones \cite{kim2021towards} and the content of titles have a significant impact on what the reader understands and remembers about a visualization \cite{borkin2015beyond}.
Additionally, one may also consider that providing the intended message of a visualization may also change the way that the reader engages with the visualization. 
For example, readers might be less inclined to spend time exploring the visualization if they are explicitly told the key insight. 
At worst, this might result in a reader skipping the visualization entirely, as has been experimentally observed with students \cite{ryoo2014designing}. 

Utilizing a tutorial about how to understand the visualization suffers from different challenges than describing the intended message because it tells the reader \textit{how} to see something instead of \textit{what} to see.
While a lack of guidance might be desirable when exploration of the data is important, it is much less helpful for readers who are less experienced in drawing conclusions from what they are seeing or lack the time to sit with the visualization long enough to parse it.

\subsection{Metadata about People}
\label{section:People}

Metadata does not only refer to information about the data, but also includes information about the people involved in the process of visualization creation. 
We have identified two groups of people which have an impact on a visualization: the creators who make choices about what and how to visualize and the intended audience whose wants and needs inform what choices the creators make.

\subsubsection{Creators}
\label{section:creators}
One could consider including information about the identities and lived experiences of the visualization creators. At its simplest, one could imagine that this could include pieces of information like an individual's name, job title, gender, or country where they live. It could also include information about the creator's motivation to create the visualization or anything that the authors deem relevant to convey. For example, depending on the topic of the visualization, one might consider including information about their political affiliation or their relationship to wealth, sexuality, or privilege. 
It may also be pertinent to think broadly about who qualifies as a ``creator'' of a visualization. 
For example, if a visualization was produced on behalf of an organization, one could also imagine disclosing the identities of the organization's leaders or its management structure, as they can reflect the priorities of that organization and how much they value the perspectives of minoritized groups \cite{dataFeminism}.
Additionally, information about the funding organizations may also help answer questions about who owns the visualization or how it can be used.

\vspace{4pt}\noindent\textbf{Pros:} Sharing aspects of the authors' identities may increase transparency and lend credibility to both the individuals who created the visualization and the visualization itself \cite{munzel2016assisting}. 
This is consistent with the feedback-loops present in data provenance and trustworthiness theories where the trustworthiness of data is reflective of the trustworthiness of the provider and the trustworthiness of the provider is reflective of the trustworthiness of the data (e.g., \cite{dai2008approach}).
Further, acknowledging that a visualization is ultimately constructed by people recognizes their unseen labor \cite{d2016feminist, correll2019ethical} and breaks down what Donna Haraway calls the ``god trick'' --- where a visualization appears to ``[see] everything from nowhere'' \cite{haraway1988situated} and thus appears to represent objective, global truth rather than a situated viewpoint.
Data products like data visualizations only ``seem objective... because the perspectives of those who produce them... pass for the default'' when really they represent one of many perspectives \cite{dataFeminism}.
This also means that providing information about the identities of the creator(s) allows the reader to more clearly see who is involved in the creation of data visualizations and to uplift the voices and work of authors in minoritized groups.

\vspace{4pt}\noindent\textbf{Cons:} 
On the other hand, if visualization readers feel that they have some reason to distrust the individual or organization who produced the visualization, they may distrust the visualization too. While in some situations this loss of trust might be considered warranted, it is also important to recognize that trust is socially constructed and that what/who is considered trustworthy is a reflection of existing power structures. This means that individuals in minoritized groups may not be trusted or believed even when they are experts -- a phenomenon commonly referred to as testimonial injustice \cite{lanius2015fact, fricker2007epistemic}. 
It is not yet clear how disbelief of an individual or institution transfers to the visualizations that they create, though recent work on visualization recommendations found that reader's trust was directly impacted by how capable they thought the creator was of making accurate visualizations \cite{zehrung2021vis}.
Unfortunately, this suggests that individuals from minoritized groups may be disproportionately hurt by the disclosure of information about identity.
However, there is evidence that readers make assumptions and judgements about the identities held by authors even when it is not disclosed (e.g., using the subject or content of a paper \cite{king2018systematic}), which suggests that witholding identity information is not an effective way to prevent the effect of harmful biases such as racism, xenophobia, or misogyny.

\subsubsection{Intended Audience}
In addition to the creators of a visualization, it may also be fruitful to consider \textit{for whom} the visualization was made. One could imagine identifying the intended audience through demographic features (e.g., Georgia residents) or by specifying a group who use visualizations in a particular way (e.g., casual visualization readers). 

\vspace{4pt}\noindent\textbf{Pros:} One benefit of disclosing the intended audience of a visualization is formally recognizing that every design does not work equally well for every reader. 
By making the intended audience of a visualization explicit, readers might get a better understanding of why particular design choices were made. For example, explaining that a map of sea temperature was intended for use by museum visitors might help a scientist understand why a red-blue scale, which maps to customary ideas of hot and cold, was used instead of the rainbow scale commonly used by experts in the field (example from \cite{phipps2010seeing}).

\vspace{4pt}\noindent\textbf{Cons:} Nonetheless, making the intended audience of a visualization explicit may also alienate some readers. For example, if a reader belongs to the intended group and does not understand what is being shown or otherwise does not find the visualization helpful, they may disengage or feel disillusioned.
Alternately, if a reader does not belong to the intended group, they may feel as if they cannot or should not try to understand what is being shown. It is possible that this could lead to disengaging entirely with a visualization. For example, people who felt that particular visualizations were too hard for them, but not too hard for others, were resistant to spend time with the visualization \cite{kennedy2016engaging}.

\section{A few open research questions}
Based on our exploration of the potential benefits and challenges of providing metadata, it is clear that there are many unanswered questions about what kind of metadata to provide and how to do so effectively. We would like to highlight a few of these open questions that we hope the visualization research community will investigate.

\vspace{4pt}\noindent\textbf{Making Metadata Usable} 
For metadata to be helpful to the reader, it needs to be presented in ways that are understandable and answers questions that the reader actually has. What information would readers find the most helpful? How does this change with the role or persona of the reader? How does this change with the purpose of a visualization (e.g., communication, leisure, analysis), its subject, or visual complexity? What role should visualization have in the process of communicating metadata about visualizations?

\vspace{4pt}\noindent\textbf{Incentivizing Creation} 
Ideally, the practice of providing metadata would become a standardized practice among researchers and practitioners, but it remains unclear how to incentivize this practice. Would a data visualization ``code of ethics'' that prioritizes transparency and accountability help? How could the labor of communicating metadata be distributed so that it does not fall entirely on over-worked design teams?
What would technical tools to help people collect and organize metadata look like?

\vspace{4pt}\noindent\textbf{Privacy}
Disclosing metadata, particularly about the people involved, may be a privacy concern for individuals even if disclosing that information is seen as ``good'' for society. How can we effectively communicate when metadata cannot be given? Who gets to decide what metadata gets shared and with whom? Is it fair to expect visualization creators to provide information about themselves?

\section{Conclusion}

Accompanying a visualization with metadata has the potential to help readers gain a better understanding of how a visualization came to be, its purpose, and the perspectives it represents. Unfortunately, deciding what kind of metadata to provide is difficult and requires considering trade-offs, particularly surrounding the tension between trust and transparency.
In this paper, we discussed the pros and cons of disclosing five kinds of metadata which could be provided with a data visualization
in order to help the visualization community make decisions about what metadata they could include and why.
There is work to be done to figure out exactly how to disclose and represent metadata alongside visualizations, but it is one possible step toward a more equitable future with visualizations that recognize influences, limits, and impacts of visualizations on the world.


\bibliographystyle{abbrv-doi}

\balance

\bibliography{main}
\end{document}